\documentclass[physics,article,accept,pdftex,moreauthors]{Definitions/mdpi}
\usepackage{amsmath}
\usepackage{amsfonts}
\usepackage{amssymb,bm}
\usepackage{siunitx}
\usepackage{color}
\usepackage{braket}
\usepackage{graphics}

\firstpage{1013}
\pubvolume{5}
\issuenum{4}
\articlenumber{66}
\pubyear{2023}
\copyrightyear{2023}
\datereceived{1 September 2023}
\daterevised{11 October 2023} 
\dateaccepted{19 October 2023}
\datepublished{25 October 2023 }
\hreflink{https://doi.org/10/3390/physics5040066} 

\Title{The Casimir Force between Two Graphene Sheets: 2D Fresnel Reflection
Coefficients, Contributions of Different Polarizations,
and the Role
of Evanescent Waves }

\TitleCitation{The Casimir Force between Two Graphene Sheets: 2D Fresnel Reflection
Coefficients, Contributions of Different Polarizations, 
and the Role
of Evanescent Waves }

\Author{{Galina L. Klimchitskaya}  $^{1,2}$\orcidA{}
and
Vladimir M. Mostepanenko $^{1,2,3,}$*\orcidB{}}

\AuthorNames{Galina L. Klimchitskaya,
Vladimir M. Mostepanenko}

\AuthorCitation{{Klimchitskaya, G.L.;} 
  Mostepanenko, V.M.}

\address{%
$^{1}$ \quad Central Astronomical Observatory at Pulkovo of the Russian Academy of Sciences, \mbox{196140 Saint Petersburg, Russia}; {g.klimchitskaya@gmail.com} 
\\
$^{2}$ \quad {Peter the Great Saint Petersburg
Polytechnic University,} 
 195251 Saint Petersburg, Russia\\
$^{3}$ \quad {Kazan Federal University,} 
 420008 Kazan, Russia}

\corres{Correspondence: vmostepa@gmail.com}

\abstract{We consider the Casimir pressure between two graphene sheets
and contributions to it determined by evanescent and propagating
waves with different polarizations.
For this purpose, the derivation of the 2-dimensional (2D) 
Fresnel reflection
coefficients on a graphene sheet is presented in terms of the transverse
and longitudinal dielectric permittivities of graphene with due account
of the spatial dispersion. The explicit expressions for both dielectric
permittivities as the functions of the 2D wave vector, frequency, and
temperature are written along the real frequency axis in the regions
of propagating and evanescent waves and at the pure imaginary
Matsubara frequencies using the polarization tensor of graphene. It is
shown that in the application region of the Dirac model nearly the total
value of the Casimir pressure between two graphene sheets is determined
by the electromagnetic field with transverse magnetic (TM) polarization.
By using the Lifshitz formula written along the real frequency axis,
the contributions of the TM-polarized propagating and evanescent waves
into the total pressure are determined. By confronting these results
with the analogous results found for plates made of real metals, the
way for bringing the Lifshitz theory using the realistic response
functions in agreement with measurements of the Casimir force between
metallic test bodies is pointed out.
}

\keyword{graphene; Casimir pressure; Dirac model; spatial dispersion;
polarization tensor; propagating waves; evanescent waves}

\begin{document}

\section{Introduction} \label{sec1}

By now, graphene has assumed great importance in the field of fundamental
physics and its numerous applications, where it plays a broad spectrum
of roles~\cite{1,2}. The distinctive characteristic features of
graphene, as opposed to ordinary bodies, are the 2-dimensional
(2D) crystal structure
of carbon atoms and massless quasiparticles described not by the
Schr\"{o}dinger equation, but by the Dirac equation, where the speed of light is
replaced with the much smaller Fermi velocity. As a result, at energies
below a few eV, the electrical and optical properties of graphene are
well described by the relatively simple Dirac model~\cite{1,2,3,3a,4,5}.
This enables one to investigate the main features of graphene not by
using some phenomenological approach, which is the usual practice in
condensed matter physics, but on the solid basis of thermal quantum
field theory and, more specifically, quantum electrodynamics at
nonzero temperature.

The subject of this {paper} 
is the Casimir force~\cite{6}, which acts
between any two uncharged closely spaced material bodies owing to the
zero-point and thermal fluctuations of the electromagnetic field. In
his original publication~\cite{6}, Casimir calculated the force acting
between two ideal metal planes kept at zero temperature. At a later
time, E. M.
Lifshitz~\cite{7,8,8a} developed the general theory expressing
the Casimir force between two plates at any temperature via the
reflection coefficients written in terms of the frequency-dependent
dielectric permittivities of plate materials. In recent years, the
Casimir force continues to grow in popularity due to the role it
plays in quantum field theory, elementary particle physics, condensed
matter, atomic physics, and even cosmology (see the monographs
\cite{9,10,11}).

Experiments measuring the Casimir force between metallic test bodies
faced problems when comparing the measurement data with theoretical
predictions of the Lifshitz theory. It turned out that if the
low-frequency response of metals is described by the universally used
dissipative Drude model, the obtained theoretical predictions are
excluded by the measurement data. If, however, the low-frequency
response is described by the dissipationless plasma model, which
should not be applicable at low frequencies, the theory gives results
in agreement with the experiment (see~\cite{10,12,12a,13,14} for a
review). Quite recently, it was shown~\cite{15} that the roots of the
problem are not in the Drude model as a whole, but only in its
possible incorrectness in the restricted area of transverse electric
evanescent waves where it has no sufficient experimental confirmation.

The response functions of metals, including the Drude model, are of
more or less phenomenological character. In this regard, of special
interest is the Casimir effect in graphene systems, which has drawn
the attention of many authors. At the early stages of investigation, the
response of graphene to the electromagnetic field was also described
by means of phenomenological methods based on the 2D Drude model,
density functional theory, Boltzmann transport equation, random phase
approximation, Kubo theory, hydrodynamic model, etc., and the obtained
results were used to calculate the Casimir force in graphene systems
\cite{16,17,18,18a,19,20,21,22,23,24,25,26,27,28,29,30,31,32,32a,33,33a}.
In doing so, it was found that in the framework of the Dirac model
the spatially nonlocal response of graphene at the pure imaginary
Matsubara frequencies can be described by the polarization tensor in
(2+1)-dimensional 
space-time and calculated precisely from the first
principles of thermal quantum field theory~\cite{34,35}. These
results were generalized to the entire plane of complex frequencies
including the real frequency axis~\cite{36,37}. In such a manner,
the reflection coefficients of electromagnetic fluctuations on a
graphene sheet were expressed directly via the components of the
polarization tensor.

The results of first-principles calculations of the Casimir force
between two graphene sheets using the polarization tensor were
compared~\cite{38} with those obtained using various phenomenological
methods, and serious limitations of the latter were demonstrated.
What is more, the measurement data of two experiments measuring
the Casimir force in graphene systems were compared with the
predictions of the Lifshitz theory using the reflection coefficients
on graphene expressed via the polarization tensor and found to be in
excellent agreement~\cite{39,40,41,42}. {{Specifically, the most precise
measurements}}~\cite{41,42} {{confirmed the theoretical prediction of}}~\cite{20} {{that for graphene systems a big thermal effect in the Casimir
force arises at much shorter separations than for metallic or dielectric bodies.}}

Thus, in the case of graphene, the Lifshitz theory does not suffer
from a problem arising for metallic plates whose electromagnetic
response was determined on partially phenomenological grounds
(we recall that the experimental data for the complex index of
refraction of metals are available only in the frequency region
above some minimum frequency and are usually extrapolated by the
Drude model to below this frequency~\cite{43}). One can conclude that
graphene supplies us with some kind of road map on how to correctly
describe the Casimir force between metallic plates. Because of this,
it is important to compare both theoretical descriptions in parallel,
including the form of reflection coefficients, the contributions of
different polarizations of the electromagnetic field, and the
propagating and evanescent waves.

In the current study,
we underline that the reflection coefficients on
a graphene sheet expressed via the polarization tensor are nothing
more than the 2D Fresnel reflection coefficients expressed via the
spatially nonlocal longitudinal and transverse dielectric
permittivities. It is stressed that for a 2D graphene sheet, as
opposed to the 3D Casimir configurations, the spatial dispersion
can be taken into account exactly on a rigorous theoretical basis.
Then, it is shown that in the application region of the Dirac model
the Casimir force between two pristine graphene sheets is completely
determined by the transverse magnetic polarization of the
electromagnetic field. In doing so, at short separations up to
hundreds of nanometers, both the propagating and evanescent waves
make essential contributions to the Casimir force, whereas at larger
separations the total force value is mostly determined by the
contribution of evanescent waves. This is compared with the case of
metallic plates where the evanescent waves play an important role in
the problem of disagreement between the predictions of the Lifshitz
theory using the Drude model and the measurement data.

The paper
is organized as follows.
 In Section~\ref{sec2}, we present the
detailed derivation of the 2D Fresnel reflection coefficients on
a graphene sheet in terms of the dielectric permittivities of
graphene with an accurate account of the spatial dispersion.
In Section~\ref{sec3}, the explicit expressions for the transverse and
longitudinal dielectric permittivities of graphene at any
temperature are presented. Section~\ref{sec4} contains the results of the
numerical computations of contributions to the Casimir pressure
between two graphene sheets due to different polarizations of the
propagating and evanescent waves. 
Section~\ref{sec5}, 
 provides 
a discussion of both similarities and distinctions in the Casimir
pressures between metallic plates and graphene sheets. Finally,
Section~\ref{sec6} contains our conclusions.

\section{Fresnel Reflection Coefficients on a Two-Dimensional Sheet} \label{sec2}
\newcommand{\ve}{{\varepsilon}}
\newcommand{\kb}{{\mbox{\boldmath$k$}}}
\newcommand{\qb}{{\mbox{\boldmath$q$}}}
\newcommand{\ob}{{\mbox{\boldmath$\rho$}}}
\newcommand{\nb}{{\mbox{\boldmath$n$}}}
\newcommand{\rb}{{\mbox{\boldmath$r$}}}
\newcommand{\eb}{{\mbox{\boldmath$E$}}}
\newcommand{\hb}{{\mbox{\boldmath$H$}}}
\newcommand{\db}{{\mbox{\boldmath$D$}}}
\newcommand{\bb}{{\mbox{\boldmath$B$}}}
\newcommand{\jb}{{\mbox{\boldmath$j$}}}
\newcommand{\sg}{{\sigma^{\rm 2D}}}
\newcommand{\lsg}{{\sigma^{\rm 2D,L}}}
\newcommand{\tsg}{{\sigma^{\rm 2D,Tr}}}

It is 
known 
that the Casimir force between two parallel plates is
expressed by the Lifshitz formula through the amplitude reflection coefficients
of the electromagnetic waves. For the ordinary three-dimensional plates, these
are the familiar Fresnel reflection coefficients written in terms of the
frequency-dependent dielectric permittivity of the voluminous plate material.
Graphene is a two-dimensional sheet of carbon atoms. Its dielectric permittivity
is spatially nonlocal and essentially depends not only on the frequency, but on
the wave vector and also on temperature.   The expressions for the
two-dimensional analogues of the Fresnel coefficients presented in terms of the
dielectric permittivity of a 2D material are not 
quite
known (see, 
e.g.,
~\cite{44,45,46}, where the transverse magnetic coefficient
\cite{44,45} and
both reflection coefficients~\cite{46} were expressed in terms of the 2D
conductivity with no account of spatial dispersion, or~\cite{11,27} where they are
presented with no detailed derivation).

Below, we  demonstrate in detail that the reflection coefficients on a 2D sheet
are obtainable in close analogy to the standard 3D Fresnel reflection coefficients,
but with due account of the spatial dispersion.

\begin{figure}[H]
\includegraphics[width=4.in]{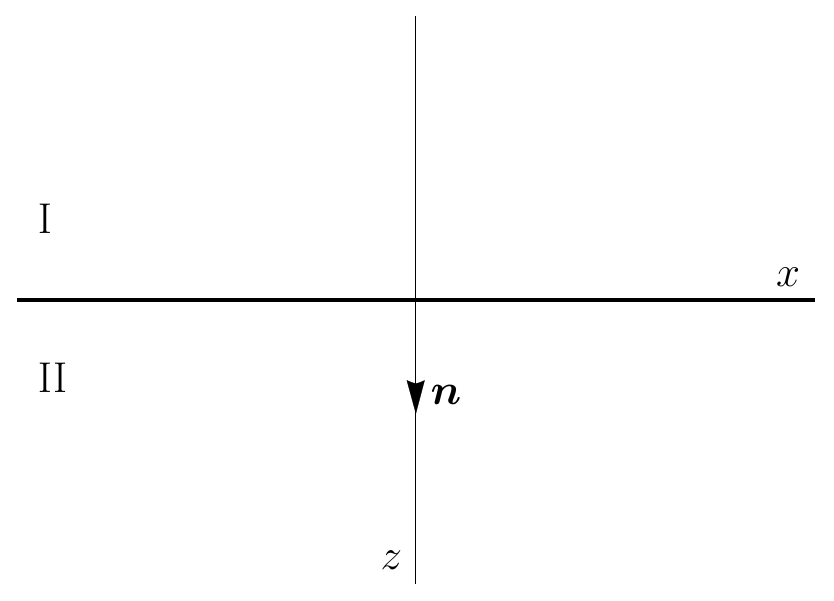}
\caption{The configuration of  a graphene sheet located at the  plane $(x,y)$ perpendicular
to the plane of the figure. The $y$-axis is directed upward. The unit normal vector
$\nb$ is directed from the region I to II along the positive direction of the $z$-axis.}
\label{fg1}
\end{figure}

Let the graphene sheet be in the plane $z=0$, where the $z$-axis is directed
downward in the plane of Figure~\ref{fg1} and the $y$-axis is directed upward
perpendicular to it.
\vspace{-3pt}
There are empty half spaces I and II on both sides of the graphene sheet.
The fluctuating electromagnetic field induces some surface charge {density,
$\varrho^{\rm 2D}(\ob,t)$,} and current {density,} 
${\mbox{\boldmath$j$}}^{\rm 2D}(\ob,t)$,
on the sheet where $\ob=(x,y)$ {and $t$ denotes the time.}
Then, the Maxwell equations in the 3D space
take the form

\begin{linenomath}
\begin{eqnarray}
&&
\nabla\db(\rb,t)=4\pi\varrho^{\rm 3D}(\rb,t),
\nonumber \\
&&
\nabla\bb(\rb,t)=0,
\nonumber \\
&&
\nabla\times\eb(\rb,t)+\frac{1}{c}\frac{\partial\bb(\rb,t)}{\partial t}=0,
\nonumber \\
&&
\nabla\times\hb(\rb,t)-\frac{1}{c}\frac{\partial\db(\rb,t)}{\partial t}=
\frac{4\pi}{c}{\mbox{\boldmath$j$}}^{\rm 3D}(\rb,t),
\label{eq1}
\end{eqnarray}
\end{linenomath}
where $\rb=(x,y,z)=(\ob,z)$, {$c$ denotes the speed of light,}
$\db$ is the electric displacement, $\bb$ is the
magnetic induction, and $\eb$ and $\hb$ are the electric and magnetic fields,
respectively. The 3D charge and current densities in Equation~(\ref{eq1}) are
given by~\cite{11,44}

\begin{linenomath}
\begin{equation}
\varrho^{\rm 3D}(\rb,t)=\varrho^{\rm 2D}(\ob,t)\delta(z), \qquad
{\mbox{\boldmath$j$}}^{\rm 3D}(\rb,t)=
{\mbox{\boldmath$j$}}^{\rm 2D}(\ob,t)\delta(z).
\label{eq2}
\end{equation}
\end{linenomath}

Note that we use the Gaussian units in Equation~(\ref{eq1}) and below.
In these units, ${\mbox{\boldmath$j$}}^{\rm 3D}$ has the dimension of
$\mbox{g}^{1/2}\mbox{cm}^{-1/2}\mbox{s}^{-2}$, whereas the dimension of
${\mbox{\boldmath$j$}}^{\rm 2D}$ is
$\mbox{g}^{1/2}\mbox{cm}^{1/2}\mbox{s}^{-2}$.

The standard electrodynamic boundary conditions on the plane $z=0$ are given by

\begin{linenomath}
\begin{eqnarray}
&&
[\db_{\rm II}(\ob,0,t)-\db_{\rm I}(\ob,0,t)]\cdot\nb=4\pi\rho^{\rm 2D}(\ob,t),
\nonumber \\
&&
[\bb_{\rm II}(\ob,0,t)-\bb_{\rm I}(\ob,0,t)]\cdot\nb=0,
\nonumber \\
&&
[\eb_{\rm II}(\ob,0,t)-\eb_{\rm I}(\ob,0,t)]\times\nb=0,
\nonumber \\
&&
[\hb_{\rm II}(\ob,0,t)-\hb_{\rm I}(\ob,0,t)]\times\nb=
-\frac{4\pi}{c}{\mbox{\boldmath$j$}}^{\rm 2D}(\ob,t),
\label{eq3}
\end{eqnarray}
\end{linenomath}
where $\nb=(0,0,1)$ is the unit vector directed along the $z$-axis
(see Figure~\ref{fg1}).

Below we assume that all fields have the form of monochromatic  plane
waves, e.g.,
\begin{linenomath}
\begin{equation}
\eb(\rb,t)=\eb^0e^{i(\kb\rb-\omega t)},\qquad
\hb(\rb,t)=\hb^0e^{i(\kb\rb-\omega t)},\qquad
\bb(\rb,t)=\bb^0e^{i(\kb\rb-\omega t)}.
\label{eq4}
\end{equation}
\end{linenomath}

Here, $\eb^0$, $\hb^0$, and $\bb^0$ are the amplitudes;
$\kb=(k_x,k_y,k_z)\equiv(\qb,k_z)$ is the 3D wave vector, and $\omega$ is
the wave frequency.

For a derivation of the Fresnel reflection coefficients on a 2D sheet, it
is
suffice to restrict our consideration to the third line of the
Maxwell 
equations 
(\ref{eq1}) and the third and fours lines in the
boundary conditions (\ref{eq3}).

Substituting Equation (\ref{eq4}) into 
the third line 
of Equations (\ref{eq1})
and (\ref{eq3}), it is readily seen that in both regions I and II

\begin{linenomath}
\begin{equation}
\kb\times\eb^0-\frac{\omega}{c}\bb^0=0
\label{eq5}
\end{equation}
\end{linenomath}
and

\begin{linenomath}
\begin{equation}
(\eb_{\rm II}^0-\eb_{\rm I}^0)\times\nb=0,
\label{eq6}
\end{equation}
\end{linenomath}
where $\eb_{\rm I}^0$ and $\eb_{\rm II}^0$ are the field amplitudes in the
regions I and II, respectively.

Now we look at the fourth line in the boundary conditions (\ref{eq3}).
Taking into account that the graphene sheet is a spatially nonlocal material,
the 2D current density in  the fourth line of
Equation (\ref{eq3}) takes the form

\begin{linenomath}
\begin{equation}
\jb^{\rm 2D}(\ob,t)=\int\limits_{-\infty}^{t}dt^{\prime}\int d^2\ob^{\prime}
\sg(\ob-\ob^{\prime},t-t^{\prime})\eb_{\rm lat}(\ob^{\prime},t^{\prime}).
\label{eq7}
\end{equation}
\end{linenomath}

Here, $\sg(\ob,t)$ is the 2D conductivity of a graphene sheet (it has the
dimension cm/s) and $\eb_{\rm lat}$ is the projection of the electric field
on the plane of graphene calculated at $z=0$: 

\begin{linenomath}
\begin{equation}
\eb_{\rm lat}(\ob,t)=\eb(\ob,0,t)-\nb\left(\eb(\ob,0,n)\cdot\nb\right)=
\nb\times[\eb(\ob,0,t)\times\nb].
\label{eq8}
\end{equation}
\end{linenomath}

Substituting Equations (\ref{eq4}) and (\ref{eq7}) into the fourth line of
Equation (\ref{eq3}), one obtains

\begin{linenomath}
\begin{equation}
(\hb_{\rm II}^0-\hb_{\rm I}^0)\times\nb=-\frac{4\pi}{c}
\sg(\qb,\omega)\eb_{\rm lat}^0,
\label{eq9}
\end{equation}
\end{linenomath}
where $\sg(\qb,\omega)$   is the Fourier image of $\sg(\ob,t)$ in the 2D
space and time, $\qb$ is the 2D wave vector, and $\eb_{\rm lat}^0$ is the
amplitude of the quantity (\ref{eq8}):
\begin{linenomath}
\begin{eqnarray}
&&
\eb_{\rm lat}(\ob,t)=\eb_{\rm lat}^0 e^{i(\qb\ob-\omega t)},
\nonumber \\
&&
\eb_{\rm lat}^0=\nb\times[\eb^0\times\nb].
\label{eq10}
\end{eqnarray}
\end{linenomath}

Note that by introducing $\sg(\qb,\omega)$ we have used the translational
invariance in the plane of a graphene sheet. In the standard Casimir problems,
where the plates are made of 3D materials separated by a gap, there is no
translational invariance in the 3D space and it is impossible to rigorously
introduce the conductivity $\sigma^{\rm 3D}(\kb,\omega)$ (and the dielectric
permittivity) depending on the 3D vector $\kb$. Because of this, for taking
into account the effects of spatial dispersion, it is necessary to use some
approximations, such as the suggestion of specular reflection
\cite{47,48}.

We recall also that the spatially dispersive materials are characterized by
the two independent conductivities, in our case, $\lsg(\qb,\omega)$ and
$\tsg(\qb,\omega)$, depending on whether $\eb_{\rm lat}^0$ in Equation (\ref{eq9})
is parallel or perpendicular to the wave vector $\qb$, respectively~\cite{49,50}.
These conductivities 
are called the longitudinal and transverse ones.

We are coming now to the derivation of the amplitude reflection coefficients on
a graphene sheet for two independent polarizations of the electromagnetic field
using Equations (\ref{eq5}), (\ref{eq6}), and (\ref{eq9}).

\begin{figure}[H]
\includegraphics[width=4.in]{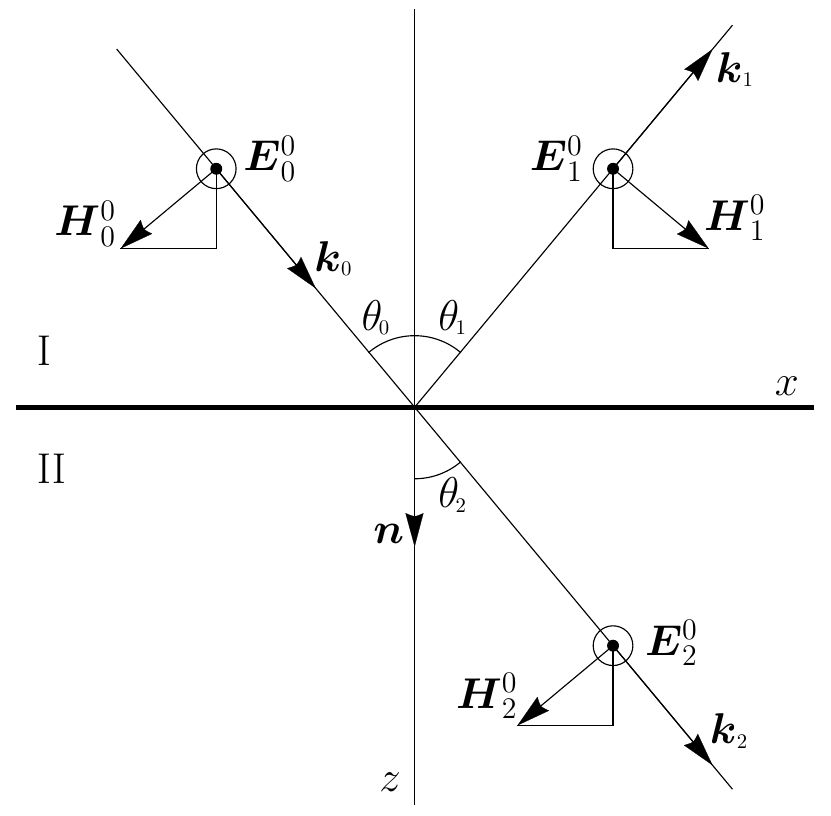}
\caption{The electromagnetic wave with the transverse electric polarization is incident on
a graphene sheet. The amplitudes of the incident, $\eb_0^0$, reflected, $\eb_1^0$,
and transmitted, $\eb_2^0$, electric field are perpendicular to the plane of
incidence and directed in the positive direction of the $y$-axis perpendicular to the
plane of the figure. The corresponding amplitudes of the magnetic field, $\hb_0^0$,
$\hb_1^0$, and $\hb_2^0$, lie in the plane of incidence, whereas $\kb_0$, $\kb_1$,
and $\kb_2$ are the corresponding wave vectors.}
\label{fg2}
\end{figure}

Let us start with the case of transverse electric polarization when the
amplitudes of the electric field of the incident, $\eb_0^0$, transmitted, $\eb_2^0$,
and reflected, $\eb_1^0$,  waves are perpendicular to the plane of incidence
$(x,z)$ and directed along the positive direction of the $y$-axis
(see Figure~\ref{fg2}). The corresponding wave vectors are $\kb_0$, $\kb_2$, and
$\kb_1$, and the amplitudes of the magnetic field, which lie in the plane of
incidence,  are $\hb_0^0$, $\hb_2^0$, and $\hb_1^0$.

Taking into account that the 2D sheet spaced in the plane $(x,y)$ or,
equivalently, $z=0$ is spatially homogeneous, one finds $k_{0x}=k_{1x}=k_{2x}$.
Considering also that $k_0^2=k_1^2=k_2^2=\omega^2/c^2$ because the space
outside of a graphene sheet is empty, 
one obtains 

\begin{linenomath}
\begin{equation}
\sin\theta_0=\frac{k_{0x}}{k_0}=\sin\theta_1=\frac{k_{1x}}{k_1}=
\sin\theta_2=\frac{k_{2x}}{k_2},
\label{eq11}
\end{equation}
\end{linenomath}
i.e., in our case, all the three angles are  equal.

According to Figure~\ref{fg2},

\begin{linenomath}
\begin{equation}
\eb_{\rm I}^0=\eb_0^0+\eb_1^0, \qquad \eb_{\rm II}^0=\eb_2^0,
\label{eq12}
\end{equation}
\end{linenomath}
where $\eb_0^0=(0,E_{0y}^0,0)$, $\eb_1^0=(0,E_{1y}^0,0)$, and
$\eb_2^0=(0,E_{2y}^0,0)$.

Taking this into account, the boundary condition (\ref{eq6}) reduces to

\begin{linenomath}
\begin{equation}
E_{0y}^0+E_{1y}^0=E_{2y}^0.
\label{eq13}
\end{equation}
\end{linenomath}

\vspace{-6pt}

The boundary condition (\ref{eq9}), where $\eb_{\rm lat}^0$ is defined  in
Equation~(\ref{eq10}), is more complicated. In view of Equation (\ref{eq6}),
both $\eb_{\rm I}^0$ and $\eb_{\rm II}^0$ can be substituted into
Equation (\ref{eq10}) in place of $\eb^0$. We choose $\eb_{\rm II}^0$ for the
sake of brevity. Then the condition (\ref{eq9}) takes the form

\begin{linenomath}
\begin{equation}
(\hb_{\rm II}^0-\hb_{\rm I}^0)\times\nb=-\frac{4\pi}{c}\tsg(\qb,\omega)
[\nb\times[\eb_{\rm II}^0\times\nb]].
\label{eq14}
\end{equation}
\end{linenomath}

Here, we took into account that the electric field is perpendicular to $\qb$.

{}From Figure~\ref{fg2}, it follows that

\begin{linenomath}
\begin{equation}
\hb_{\rm I}^0=\hb_0^0+\hb_1^0, \qquad \hb_{\rm II}^0=\hb_2^0,
\label{eq15}
\end{equation}
\end{linenomath}
where $\hb_0^0=(H_{0x}^0,0,H_{0z}^0)$, $\hb_1^0=(H_{1x}^0,0,H_{1z}^0)$, and
$\hb_2^0=(H_{2x}^0,0,H_{2z}^0)$.

Substituting Equations (\ref{eq12}) and (\ref{eq15}) into the boundary condition
(\ref{eq14}),  one obtains 
after the elementary algebra:

\begin{linenomath}
\begin{equation}
H_{2x}^0-H_{0x}^0-H_{1x}^0=\frac{4\pi}{c}\tsg(\qb,\omega)E_{2y}^0.
\label{eq16}
\end{equation}
\end{linenomath}

From the Maxwell equation 
(\ref{eq5}), written for the incident wave in free
space where $\bb^0=\hb^0$, one obtains

\begin{linenomath}
\begin{equation}
\hb_0^0=\frac{c}{\omega}[\kb_0\times\eb_0^0].
\label{eq17}
\end{equation}
\end{linenomath}

With account of $\kb_0=(k_{0x},0,k_{0z})$ and $\eb_0^0=(0,E_{0y}^0,0)$,
this reduces to

\begin{linenomath}
\begin{equation}
H_{0x}^0=-\frac{c}{\omega}k_{0z}E_{0y}^0,
\label{eq18}
\end{equation}
\end{linenomath}
where, considering Equation (\ref{eq11}) and using that in this case $k_{0x}$ plays
the role of $q$,

\begin{linenomath}
\begin{equation}
k_{0z}=\frac{\omega}{c}\cos\theta_0=\sqrt{\frac{\omega^2}{c^2}-q^2}.
\label{eq19}
\end{equation}
\end{linenomath}

In a similar way, from Equation (\ref{eq5}) written for the reflected and
transmitted waves, one finds 
\begin{linenomath}
\begin{equation}
H_{1x}^0=-\frac{c}{\omega}k_{0z}E_{1y}^0, \qquad
H_{2x}^0=-\frac{c}{\omega}k_{0z}E_{2y}^0.
\label{eq20}
\end{equation}
\end{linenomath}

Substituting Equations (\ref{eq18}) and (\ref{eq20}) into Equation (\ref{eq16}),
we finally obtain
\begin{linenomath}
\begin{equation}
E_{0y}^0-E_{1y}^0-E_{2y}^0=\frac{4\pi\omega}{c^2k_{0z}}\tsg(\qb,\omega)E_{2y}^0.
\label{eq21}
\end{equation}
\end{linenomath}

By solving this equation together with Equation (\ref{eq13}), the transverse
electric (TE) 
reflection coefficient is found in the form

\begin{linenomath}
\begin{equation}
r_{\rm TE}(\qb,\omega)=\frac{E_{1y}^0}{E_{0y}^0}=
-\frac{2\pi\omega\tsg(\qb,\omega)}{c^2k_{0z}+2\pi\omega\tsg(\qb,\omega)}.
\label{eq22}
\end{equation}
\end{linenomath}

Note that in~Ref. \cite{46} this reflection coefficient was obtained in the special
case of normal incidence with ignored spatial dispersion.

By taking into account that for a 2D sheet, the spatially nonlocal dielectric
permittivity is expressed via the conductivity as~\cite{11,27}

\begin{linenomath}
\begin{equation}
\ve^{\rm 2D,Tr}(\qb,\omega)=1+\frac{2\pi i\tsg(\qb,\omega)q}{\omega},
\label{eq23}
\end{equation}
\end{linenomath}
and using Equation (\ref{eq19}), we rewrite the reflection coefficient (\ref{eq22})
in the final form

\begin{linenomath}
\begin{equation}
r_{\rm TE}(\qb,\omega)=-\frac{\omega^2[\ve^{\rm 2D,Tr}(\qb,\omega)-
1]}{ic^2q\sqrt{\frac{\omega^2}{c^2}-q^2}+\omega^2
[\ve^{\rm 2D,Tr}(\qb,\omega)-1]}{}.
\label{eq24}
\end{equation}
\end{linenomath}

This is the transverse electric Fresnel reflection coefficient on a 2D graphene
sheet expressed via the spatially nonlocal transverse dielectric permittivity
of graphene.

\begin{figure}[H]
\includegraphics[width=4.in]{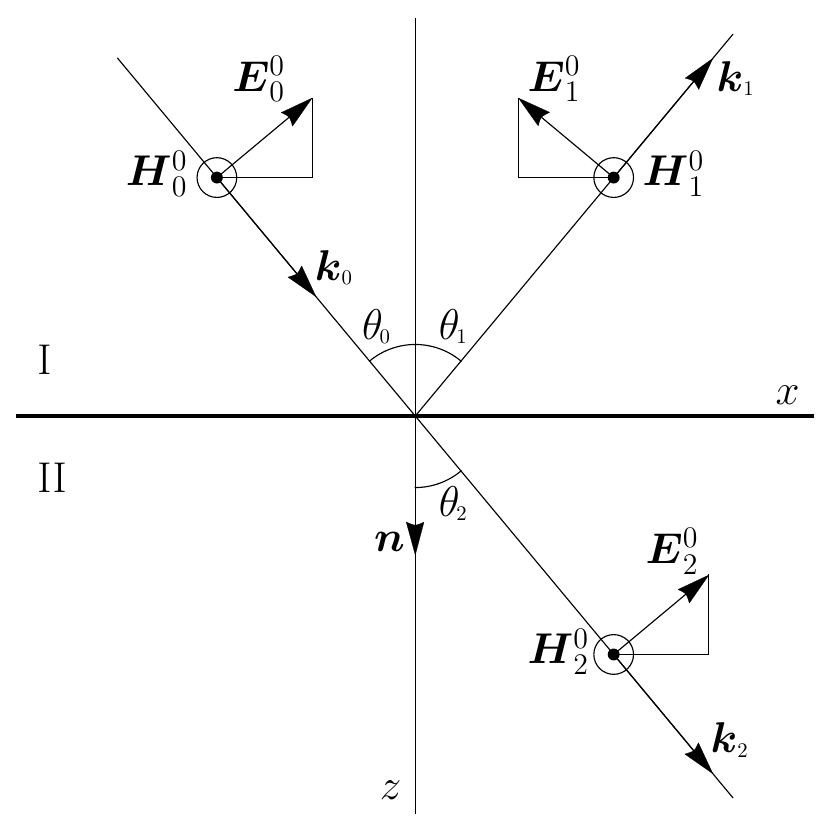}
\caption{The electromagnetic wave with the transverse magnetic polarization is incident on
a graphene sheet. The amplitudes of the incident, $\hb_0^0$, reflected, $\hb_1^0$,
and transmitted, $\hb_2^0$, magnetic field are perpendicular to the plane of
incidence and directed in the positive direction of the $y$-axis perpendicular to the
plane of the figure. The corresponding amplitudes of the electric field, $\eb_0^0$,
$\eb_1^0$, and $\eb_2^0$, lie in the plane of incidence, whereas $\kb_0$, $\kb_1$,
and $\kb_2$ are the corresponding wave vectors.
}
\label{fg3}
\end{figure}

We now proceed to a derivation of the transverse magnetic reflection coefficient
on a graphene sheet. In this case, the amplitudes of the magnetic field of the
incident, $\hb_0^0$,  transmitted, $\hb_2^0$, and reflected, $\hb_1^0$, waves
are perpendicular to the plane of incidence and directed along the positive
direction of the $y$-axis (see Figure~\ref{fg3}). The amplitudes of the electric
field, $\eb_0^0$,  $\eb_2^0$, and $\eb_1^0$, lie in the plane of incidence.

According to Figure~\ref{fg3},

\begin{linenomath}
\begin{equation}
\hb_{\rm I}^0=\hb_0^0+\hb_1^0, \qquad \hb_{\rm II}^0=\hb_2^0,
\label{eq25}
\end{equation}
\end{linenomath}
where $\hb_0^0=(0,H_{0y}^0,0)$, $\hb_1^0=(0,H_{1y}^0,0)$, and
$\hb_2^0=(0,H_{2y}^0,0)$.

Taking into account that, in this case,

\begin{linenomath}
\begin{equation}
E_{\rm II}^0=\eb_2^0=(E_{2x}^0,0,E_{2z}^0),
\label{eq26}
\end{equation}
\end{linenomath}
one obtains

\begin{linenomath}
\begin{eqnarray}
&&
[\eb_2^0\times\nb]=(0,-E_{2x}^0,0),
\nonumber \\
&&
\nb\times[\eb_2^0\times\nb]=(E_{2x}^0,0,0)=
(E_2^0\cos\theta_0,0,0),
\label{eq27}
\end{eqnarray}
\end{linenomath}
where $\cos\theta_0$ is defined in Equation (\ref{eq19}).

The boundary condition (\ref{eq9}), (\ref{eq10}) takes the form

\begin{linenomath}
\begin{equation}
(\hb_{\rm II}^0-\hb_{\rm I}^0)\times\nb=-\frac{4\pi}{c}
\lsg(\qb,\omega)[\nb\times[E_{\rm II}^0\times\nb]]
\label{eq28}
\end{equation}
\end{linenomath}

The longitudinal conductivity $\lsg$ appears in this equation because the
2D wave vector $\qb$ is now parallel to $\eb_{\rm lat}^0$.

Substituting Equations (\ref{eq25}) and (\ref{eq27}) into the boundary
condition (\ref{eq28}), one finds 

\begin{linenomath}
\begin{equation}
H_{2y}^0-H_{0y}^0-H_{1y}^0=-\frac{4\pi\lsg(\qb,\omega)}{c}E_2^0\cos\theta_0.
\label{eq28a}
\end{equation}
\end{linenomath}

For the transverse magnetic polarization, the boundary condition (\ref{eq6})
reduces to

\begin{linenomath}
\begin{equation}
E_{0x}^0+E_{1x}^0-E_{2x}^0=0.
\label{eq29}
\end{equation}
\end{linenomath}

With account of Equation (\ref{eq11}), which is valid for both polarizations
of the electromagnetic field, 
Equation~(\ref{eq29}) 
is equivalent to

\begin{linenomath}
\begin{equation}
E_0^0\cos\theta_0-E_1^0\cos\theta_0-E_2^0\cos\theta_0=0
\label{eq30}
\end{equation}
\end{linenomath}
and finally to

\begin{linenomath}
\begin{equation}
E_0^0-E_1^0-E_2^0=0.
\label{eq31}
\end{equation}
\end{linenomath}

Let us now use the Maxwell equation 
(\ref{eq5}) for the incident wave
$\hb_0^0=\bb_0^0=(0,H_{0y}^0,0)$.  Then it takes the  form of
Equation (\ref{eq17}). By using $\eb_0^0=(E_{0x}^0,0,E_{0z}^0)$, one
obtains from Equation (\ref{eq17}) with the help of Equations (\ref{eq11})
and (\ref{eq19})

\begin{linenomath}
\begin{equation}
H_{0y}^0=\frac{c}{\omega}(k_{0z}E_{0x}^0-k_{0x}E_{0z}^0)=
\frac{c}{\omega}(k_{0z}\cos\theta_0+k_{0x}\sin\theta_0)E_0^0=E_0^0.
\label{eq32}
\end{equation}
\end{linenomath}

In a similar way, from the Maxwell
equation
(\ref{eq17}) applied to the
reflected and transmitted waves, one obtains

\begin{linenomath}
\begin{equation}
H_{1y}^0=E_1^0, \qquad H_{2y}^0=E_2^0.
\label{eq33}
\end{equation}
\end{linenomath}

Substituting Equations (\ref{eq32}) and (\ref{eq33}) into Equations (\ref{eq28a})
and (\ref{eq31}), one finds 

\begin{linenomath}
\begin{eqnarray}
&&
H_{2y}^0-H_{0y}^0-H_{1y}^0=-\frac{4\pi\lsg(\qb,\omega)}{c}H_{2y}^0\cos\theta_0,
\nonumber \\
&&
H_{0y}^0-H_{1y}^0-H_{2y}^0=0.
\label{eq34}
\end{eqnarray}
\end{linenomath}

By solving these equations together, we derive the transverse magnetic (TM)
reflection coefficient on a 2D graphene sheet:

\begin{linenomath}
\begin{equation}
r_{\rm TM}(\qb,\omega)=\frac{H_{1y}^0}{H_{0y}^0}=
\frac{2\pi\lsg(\qb,\omega)\cos\theta_0}{c+2\pi\lsg(\qb,\omega)\cos\theta_0}.
\label{eq35}
\end{equation}
\end{linenomath}

The result (\ref{eq35}) was obtained
in~Refs. \cite{44,45,46} in the spatially local case. In the presence
of spatial dispersion, both the results (\ref{eq22}) and (\ref{eq35}) are
presented in~Refs. \cite{11,27} with no detailed derivation.

By  using an expression for the longitudinal dielectric
permittivity of a 2D sheet through its conductivity~\cite{11,27}

\begin{linenomath}
\begin{equation}
\ve^{\rm 2D,L}(\qb,\omega)=1+\frac{2\pi i\lsg(\qb,\omega)q}{\omega},
\label{eq36}
\end{equation}
\end{linenomath}
and Equation (\ref{eq19}), one obtains

\begin{linenomath}
\begin{equation}
r_{\rm TM}(\qb,\omega)=\frac{[\ve^{\rm 2D,L}(\qb,\omega)-
1]\sqrt{\frac{\omega^2}{c^2}-q^2}}{iq+
[\ve^{\rm 2D,L}(\qb,\omega)-1]\sqrt{\frac{\omega^2}{c^2}-q^2}}.
\label{eq37}
\end{equation}
\end{linenomath}

This is the transverse magnetic Fresnel reflection coefficient on a 2D graphene
sheet expressed via the longitudinal dielectric permittivity
of graphene.

\section{Spatially Nonlocal Dielectric Permittivities of Graphene and
the Polarization Tensor} \label{sec3}
\newcommand{\wq}{{(\mbox{\boldmath$q$},\omega)}}
\newcommand{\xq}{{(\mbox{\boldmath$q$},i\xi_l)}}
\newcommand{\Lve}{\varepsilon^{\rm 2D,L}}
\newcommand{\Tve}{\varepsilon^{\rm 2D,Tr}}

It is common knowledge that at low energies (smaller than approximately 3~eV~\cite{51})
graphene is well described by the Dirac model as a set of massless quasiparticles
satisfying the Dirac equation, where the speed of light $c$ is replaced with the
Fermi velocity \mbox{$v_F\approx c/300$}~\cite{1,2,3}. In the framework of the Dirac model,
it is possible to derive explicit expressions for the polarization tensor of graphene,
which describes the response of a graphene sheet to the electromagnetic field
\cite{34,35,36,37}, and thus find both the transverse and longitudinal permittivities
of graphene starting from the first principles of quantum electrodynamics. The dielectric
permittivities obtained in this way depend on the wave vector, on the frequency, and
also on temperature.

The polarization tensor of graphene in (2+1)-dimensional space-time is notated as
$\Pi_{\mu\nu}\wq$, where $\mu,\nu=0,\,1,\,2$  and the dependence on temperature
is implied (here, we consider the pristine graphene sheet with no energy gap in the
spectrum of quasiparticles and perfect hexagonal crystal lattice).
The transverse dielectric permittivity of graphene is expressed as~\cite{52}
\begin{linenomath}
\begin{equation}
\Tve\wq-1=-\frac{c^2}{2\hbar q\omega^2}\Pi\wq,
\label{eq38}
\end{equation}
\end{linenomath}
where the quantity $\Pi$ is the following combination of the components of the
polarization tensor:
\begin{linenomath}
\begin{equation}
\Pi\wq\equiv q^2\Pi_{\mu}^{\,\mu}\wq+\left(\frac{\omega^2}{c^2}-q^2\right)
\Pi_{00}\wq,
\label{eq39}
\end{equation}
\end{linenomath}
{$\hbar$ is the reduced Planck's constant}
and the summation is made over the repeated indices.

The longitudinal dielectric permittivity of graphene is immediately expressed via
the 00 component of the polarization tensor ~\cite{52}

\begin{linenomath}
\begin{equation}
\Lve\wq-1=\frac{c^2}{2\hbar q}\Pi_{00}\wq,
\label{eq40}
\end{equation}
\end{linenomath}

The polarization tensor of graphene along the real frequency axis was obtained in Ref.
\cite{36}. It was considered for the propagating waves, which satisfy the
condition

\begin{linenomath}
\begin{equation}
q\leqslant\frac{\omega}{c},
\label{eq41}
\end{equation}
\end{linenomath}
and for the evanescent waves, which satisfy either the condition

\begin{linenomath}
\begin{equation}
\frac{\omega}{c}<q\leqslant\frac{\omega}{v_F}\approx 300\frac{\omega}{c}
\label{eq42}
\end{equation}
\end{linenomath}
(the so-called plasmonic region~\cite{53}) or the condition

\begin{linenomath}
\begin{equation}
q>\frac{\omega}{v_F}\approx 300\frac{\omega}{c}.
\label{eq43}
\end{equation}
\end{linenomath}

Using the expression 
from Ref.~\cite{36} 
for $\Pi$ in the region of propagating waves
(\ref{eq41}) and in the plasmonic region (\ref{eq42}), the transverse dielectric
permittivity of graphene (\ref{eq38}) in these regions can be written in the
same form:
\begin{linenomath}
\begin{eqnarray}
&&
\Tve\wq-1=i\pi\alpha q\frac{c}{2\omega^2}\sqrt{\omega^2-v_F^2q^2}
\nonumber \\
&&~~~~~
-\frac{8\alpha c^2}{v_F^2 q}\left\{\int\limits_{0}^{u^{(-)}}\frac{du}{e^{\beta u}+1}
\left[1-\frac{1}{2\omega^2}\sqrt{\omega^2-v_F^2q^2}\sum_{\lambda=\pm 1}
B(2cu+\lambda\omega)\right]\right.
\nonumber \\
&&~~~~~~~~~~~~~~~
\left.+\int\limits_{u^{(-)}}^{\infty}\frac{du}{e^{\beta u}+1}
\left[1-\frac{1}{2\omega^2}\sqrt{\omega^2-v_F^2q^2}\sum_{\lambda=\pm 1}\lambda
B(2cu+\lambda\omega)\right]\right\}.
\label{eq44}
\end{eqnarray}
\end{linenomath}

Here and below, it is assumed that $\omega>0$ and the following notations are
introduced:
\begin{linenomath}
\begin{equation}
u^{(-)}=\frac{1}{2c}(\omega-v_Fq),\qquad \beta=\frac{\hbar c}{k_BT}, \qquad
B(x)=\frac{x^2}{\sqrt{x^2-v_F^2q^2}},
\label{eq45}
\end{equation}
\end{linenomath}
{$\alpha$ denotes the fine structure constant,} 
$k_B$ is the Boltzmann constant and $T$ is the temperature of a graphene sheet.

In the region (\ref{eq43}), using the corresponding expression for $\Pi$~\cite{36},
one obtains
another expression for the transverse dielectric permittivity of
graphene
\begin{linenomath}
\begin{eqnarray}
&&
\Tve\wq-1=-\pi\alpha q\frac{c}{2\omega^2}\sqrt{v_F^2q^2-\omega^2}
\nonumber \\
&&~~~~~
-\frac{4\alpha c}{v_F^2 q}\sqrt{v_F^2q^2-\omega^2}
\int\limits_{0}^{\infty}\frac{dw}{e^{Dw}+1}
\left[1-\frac{1}{2}\sum_{\lambda=\pm 1}
\frac{(\sqrt{v_F^2q^2-\omega^2}w+\lambda\omega)^2}{\omega^2
\sqrt{1-w^2-\frac{2\lambda\omega w}{\sqrt{v_F^2q^2-\omega^2}}}}\right],
\label{eq46}
\end{eqnarray}
\end{linenomath}
where $D=\hbar\sqrt{v_F^2q^2-\omega^2}/(2k_BT)$.

In a similar way, using the expression 
from Ref.~\cite{36} for $\Pi_{00}$ in the region
of propagating (\ref{eq41}) and plasmonic (\ref{eq42}) wave vectors and frequencies,
one finds
the explicit form of the longitudinal dielectric permittivity of graphene
(\ref{eq40}) in these regions:
\begin{linenomath}
\begin{eqnarray}
&&
\Lve\wq-1=i\pi\alpha cq\frac{1}{2\sqrt{\omega^2-v_F^2q^2}}
\nonumber \\
&&~~~~~
+\frac{8\alpha c^2}{v_F^2 q}\left\{\int\limits_{0}^{u^{(-)}}\frac{du}{e^{\beta u}+1}
\left[1-\frac{1}{2\sqrt{\omega^2-v_F^2q^2}}\sum_{\lambda=\pm 1}
F(2cu+\lambda\omega)\right]\right.
\nonumber \\
&&~~~~~~~~~~~~~~~
\left.+\int\limits_{u^{(-)}}^{\infty}\frac{du}{e^{\beta u}+1}
\left[1-\frac{1}{2\sqrt{\omega^2-v_F^2q^2}}\sum_{\lambda=\pm 1}\lambda
F(2cu+\lambda\omega)\right]\right\},
\label{eq47}
\end{eqnarray}
\end{linenomath}
where

\begin{linenomath}
\begin{equation}
F(x)=\sqrt{x^2-v_F^2q^2}.
\label{eq48}
\end{equation}
\end{linenomath}

Using the expression of $\Pi_{00}$~\cite{36} in the region (\ref{eq43}), for
the longitudinal permittivity of graphene (\ref{eq40}) in this region, one obtains

\begin{linenomath}
\begin{eqnarray}
&&
\Lve\wq-1=\pi\alpha cq\frac{1}{2\sqrt{v_F^2q^2-\omega^2}}
\nonumber \\
&&~~~~~
+\frac{4\alpha c}{v_F^2 q}\sqrt{v_F^2q^2-\omega^2}
\int\limits_{0}^{\infty}\frac{dw}{e^{Dw}+1}
\left[1-\frac{1}{2}\sum_{\lambda=\pm 1}
\sqrt{1-w^2-\frac{2\lambda\omega w}{\sqrt{v_F^2q^2-\omega^2}}}\right].
\label{eq49}
\end{eqnarray}
\end{linenomath}

Thus, both the transverse and longitudinal dielectric permittivities of graphene
are obtained in all ranges of the wave vectors and frequencies
(\ref{eq41})--(\ref{eq43}). We emphasize that the first lines of Equations
(\ref{eq44}), (\ref{eq46}), (\ref{eq47}), and (\ref{eq49}) represent the corresponding
dielectric permittivity at zero temperature. The terms in the next lines of these
equations define the thermal correction to it found in the framework of the
Dirac model. These terms make a profound effect on the reflectivity~\cite{36}
and conductivity~\cite{54} properties of graphene, and also on the Casimir
interaction between graphene sheets~\cite{38,40,41,42}.
By construction from the polarization tensor, the obtained permittivities satisfy
the Kramers--Kronig relations. The specific form of these relations was investigated
in the spatially local limit $q\to 0$~\cite{55} and at zero temperature~\cite{56}.

For the calculation of the Casimir force in graphene systems, it is helpful to use
the reflection coefficients (\ref{eq24}) and (\ref{eq37}), as well as the dielectric
permittivities of graphene, written at the pure imaginary Matsubara frequencies
$\omega=i\xi_l=2\pi ik_BTl/\hbar$, where $l=0,\,1,\,2,\,\ldots\,$.

Substituting $\omega=i\xi_l$ into Equations  (\ref{eq24}) and (\ref{eq37}), one obtains, respectively, 
\begin{linenomath}
\begin{eqnarray}
&&
r_{\rm TE}\xq=-\frac{\xi_l^2[\Tve\xq-1]}{c^2q\sqrt{q^2+\frac{\xi_l^2}{c^2}}+
\xi_l^2[\Tve\xq-1]},
\nonumber \\
&&
r_{\rm TM}\xq=\frac{[\Tve\xq-1]\sqrt{q^2+\frac{\xi_l^2}{c^2}}}{q+
[\Tve\xq-1]\sqrt{q^2+\frac{\xi_l^2}{c^2}}}.
\label{eq50}
\end{eqnarray}
\end{linenomath}

These are the Fresnel reflection coefficients in two dimensions calculated at the
pure imaginary Matsubara frequencies. The same expressions are obtained if one
substitutes Equations (\ref{eq38}) and (\ref{eq40}) into the reflection  coefficients
derived in~Refs. \cite{34,35} directly in terms of the polarization tensor.

The spatially nonlocal dielectric permittivities of graphene along the imaginary
frequency axis are immediately obtainable from Equations (\ref{eq46}) and (\ref{eq49})
valid in the interval (\ref{eq43}) by putting $\omega=i\xi_l$. The results are
\begin{linenomath}
\begin{eqnarray}
&&
\Tve\xq-1=\pi\alpha q\frac{c}{2\xi_l^2}\sqrt{v_F^2q^2+\xi_l^2}
\nonumber \\
&&~~~~~
-\frac{4\alpha c}{v_F^2 q}\sqrt{v_F^2q^2+\xi_l^2}
\int\limits_{0}^{\infty}\frac{dw}{e^{D_lw}+1}
\left[1+\frac{1}{2}\sum_{\lambda=\pm 1}
\frac{(\sqrt{v_F^2q^2+\xi_l^2}w+i\lambda\xi_l)^2}{\xi_l^2
\sqrt{1-w^2-\frac{2i\lambda \xi_l w}{\sqrt{v_F^2q^2+\xi_l^2}}}}\right],
\nonumber\\
&&
\Lve\xq-1=\pi\alpha cq\frac{1}{2\sqrt{v_F^2q^2+\xi_l^2}}
\nonumber \\
&&~~~~~
+\frac{4\alpha c}{v_F^2 q}\sqrt{v_F^2q^2+\xi_l^2}
\int\limits_{0}^{\infty}\frac{dw}{e^{D_lw}+1}
\left[1-\frac{1}{2}\sum_{\lambda=\pm 1}
\sqrt{1-w^2-\frac{2i\lambda\xi_l w}{\sqrt{v_F^2q^2+\xi_l^2}}}\right],
\label{eq51}
\end{eqnarray}
\end{linenomath}
where now $D_l=\hbar\sqrt{v_F^2q^2+\xi_l^2}/(2k_BT)$.

These expressions are indeed 
real as it should be. The same dielectric permittivities
are obtained at once from Equations (\ref{eq38}) and (\ref{eq40}) written at
$\omega=i\xi_l$ when substituting expressions for $\Pi\xq$ and $\Pi_{00}\xq$ derived
directly along the imaginary frequency axis~\cite{57} rather than analytically continued
from the real frequency axis as it was made above.

\section{Contribution of Different Polarizations and the Role of Evanescent
Waves in the Casimir Pressure between Two Graphene Sheets} \label{sec4}

The Casimir pressure between two parallel graphene sheets at temperature $T$ separated
by distance $a$ is given by the Lifshitz formula, which can be presented in
terms of either pure imaginary Matsubara or real frequencies~\cite{7,10}. In both cases,
the total pressure is the sum of contributions from the electromagnetic waves of
TM and TE polarizations.

We begin from the representation in terms of the Matsubara frequencies

\begin{linenomath}
\begin{equation}
P(a,T)=P_{\rm TM}(a,T)+P_{\rm TE}(a,T),
\label{eq52}
\end{equation}
\end{linenomath}
where

\begin{linenomath}
\begin{equation}
P_{\rm TM,TE}(a,T)=-\frac{k_BT}{\pi}\sum_{l=0}^{\infty}{\vphantom{\sum}}^{\!\prime}
\int\limits_{0}^{\infty}dq\,q\sqrt{q^2+\frac{\xi_l^2}{c^2}}
\left[r_{\rm TM,TE}^{-2}\xq\,e^{2a\sqrt{q^2+\frac{\xi_l^2}{c^2}}}-1\right]^{-1}.
\label{eq53}
\end{equation}
\end{linenomath}

Here, the prime on the summation sign adds the factor 1/2 to the term with $l=0$,
and the reflection coefficients on a graphene sheet for both polarizations are
defined in Equation (\ref{eq50}) with the dielectric permittivities of graphene
presented in Equation (\ref{eq51}).

We performed computations of both $P_{\rm TM}$ and  $P_{\rm TE}$ in the application
region of the Dirac model, i.e., under a condition that the characteristic energy
of the Casimir force $\hbar\omega_c=\hbar c/(2a)$ should be less than 3~eV~\cite{51}.
This condition is well satisfied at $a\geqslant 200~$nm, where
$\hbar\omega_c\leqslant 0.5~$eV.

\begin{figure}[H]
\includegraphics[width=5.in]{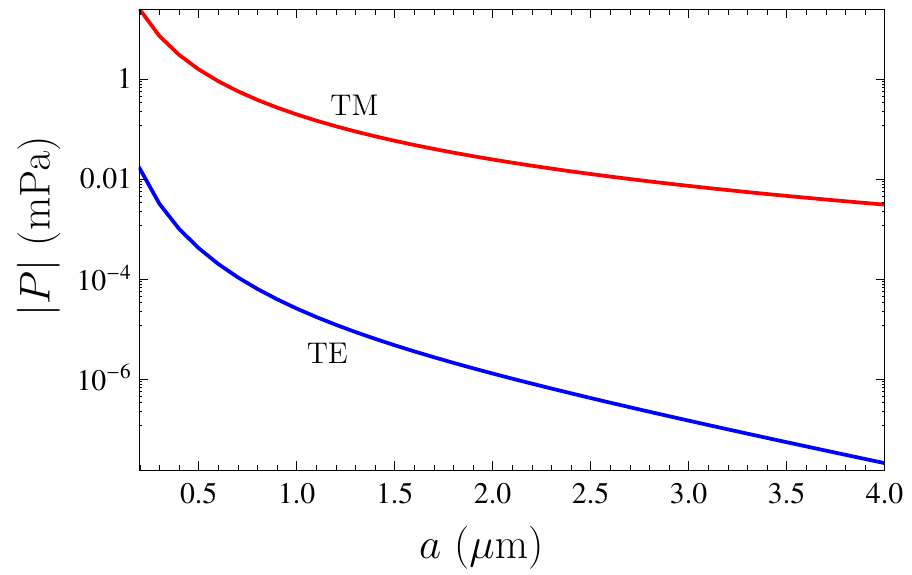}
\caption{The magnitudes of contributions of the transverse magnetic (TM) and transverse electric (TE)
polarizations to the Casimir pressure between two graphene sheets at {$T=300~$K are
shown} 
in the logarithmic scale as the function of separation by the upper
and lower lines, respectively.}
\label{fg4}
\end{figure}

The computational results for the magnitudes of $P_{\rm TM}$ and  $P_{\rm TE}$  at
$T=300~$K are presented in Figure~\ref{fg4} in the logarithmic scale by the upper
and lower lines, respectively, as the function of separation between the graphene
sheets. Both $P_{\rm TM}$ and  $P_{\rm TE}$ are negative, i.e., they contribute to the
Casimir attraction.

As seen in Figure~\ref{fg4}, the major contribution to the Casimir pressure at
$a\geqslant 200~$nm is given by the transverse magnetic polarization, whereas the
transverse electric one makes only a negligible small contribution. Thus, at
$a=200~$nm, we have $P_{\rm TM}/P_{\rm TE}=1530$, i.e., $P_{\rm TM}/P=0.99935$.
The role of the TM polarization only increases with increasing separation.
As two more examples, at $a=2$ and $4~\upmu$m one finds that
$P_{\rm TM}/P_{\rm TE}=1.92\times 10^4$ and $1.49\times 10^5$, respectively.
This results in the following respective fractions of $P_{\rm TM}$ in the total
Casimir pressure: $P_{\rm TM}/P=0.99995$ and 0.999993.

In Section~\ref{sec3}, devoted to the nonlocal dielectric permittivities of graphene, they
were considered in the region of propagating (\ref{eq41}) and evanescent
(\ref{eq42}) and (\ref{eq43}) waves. In so doing, within the region of propagating
(\ref{eq41}) and in the plasmonic subregion (\ref{eq42}) of evanescent waves, these
permittivities have a common analytic form. Nevertheless, keeping in mind an
especially important role of the propagating waves (\ref{eq41}), which are on the
mass shell in free space, it is appropriate  to consider their
contribution to the Casimir pressure separately. Then, the contribution of the
evanescent waves is computed as a sum of two terms using two different forms of
the dielectric functions depending on whether the condition (\ref{eq42}) or
(\ref{eq43}) is satisfied. Such a separation into the propagating and evanescent
waves is also dictated by the form of the Lifshitz formula written in terms of real
frequencies (see below).

The  representation mathematically equivalent to Equations (\ref{eq52}) and (\ref{eq53})
of the Lifshitz formula in terms of real frequencies can be written
in the form

\begin{linenomath}
\begin{equation}
P(a,T)=P_{\rm TM}^{\rm prop}(a,T)+P_{\rm TE}^{\rm prop}(a,T)+
P_{\rm TM}^{\rm evan}(a,T)+P_{\rm TE}^{\rm evan}(a,T).
\label{eq54}
\end{equation}
\end{linenomath}

Here, the contributions of the propagating waves with different polarizations are
given by~\cite{7,10}

\begin{linenomath}
\begin{eqnarray}
&&
P_{\rm TM,TE}^{\rm prop}(a,T)=-\frac{\hbar}{2\pi^2}\int\limits_{0}^{\infty}
d\omega\,\coth\frac{\hbar\omega}{2k_BT}\int\limits_{0}^{\omega/c}q\,dq\,
\nonumber\\
&&~~~~~~~~~~~~~\times
{\rm Im}\left\{\sqrt{q^2-\frac{\omega^2}{c^2}}\left[r_{\rm TM,TE}^{-2}(\qb,\omega)
\,e^{2a\sqrt{q^2-\frac{\omega^2}{c^2}}}-1\right]^{-1}\right\},
\label{eq55}
\end{eqnarray}
\end{linenomath}
where the reflection coefficients are defined in Equations (\ref{eq24}) and
(\ref{eq37}) and the dielectric permittivities in the region (\ref{eq41}) are
given by Equations (\ref{eq44}) and (\ref{eq47}).

The contributions of evanescent waves to Equation (\ref{eq54}) with different
polarizations take the form~\cite{7,10}

\begin{linenomath}
\begin{eqnarray}
&&
P_{\rm TM,TE}^{\rm evan}(a,T)=-\frac{\hbar}{2\pi^2}\int\limits_{0}^{\infty}
d\omega\,\coth\frac{\hbar\omega}{2k_BT}\int\limits_{\omega/c}^{\infty}q\,dq\,
\sqrt{q^2-\frac{\omega^2}{c^2}}
\nonumber\\
&&~~~~~~~~~~~~~\times
{\rm Im}\left[r_{\rm TM,TE}^{-2}(\qb,\omega)
\,e^{2a\sqrt{q^2-\frac{\omega^2}{c^2}}}-1\right]^{-1},
\label{eq56}
\end{eqnarray}
\end{linenomath}
where the reflection coefficients are again defined in Equations (\ref{eq24}) and
(\ref{eq37}). As to the dielectric permittivities entering these reflection
coefficients, in the region (\ref{eq42}), they are given by Equations (\ref{eq44})
and (\ref{eq47}), but in the region (\ref{eq43}), by Equations (\ref{eq46})
and (\ref{eq49}).

Equations (\ref{eq55}) and (\ref{eq56}) are not as convenient for computations
as Equation (\ref{eq53}). This is most pronounced in $P_{\rm TM,TE}^{\rm prop}$
defined in Equation (\ref{eq55}), which contains the quickly oscillating functions
due to the pure imaginary power in the exponential factor. As to Equation
(\ref{eq56}), the power of the exponent remains real.

\textls[-15]{Taking into account that in the application region of the Dirac model nearly
the total Casimir pressure is determined by the TM polarized waves, we compute the
quantity $P_{\rm TM}^{\rm evan}$ by Equations (\ref{eq56}) and (\ref{eq37})
using the dielectric permittivities defined in \mbox{Equations (\ref{eq47}) and
(\ref{eq49})}.} As to the  quantity $P_{\rm TM}^{\rm prop}$, it is more convenient
to not compute it directly by Equation (\ref{eq55}), but determine it as a difference

\begin{linenomath}
\begin{equation}
P_{\rm TM}^{\rm prop}(a,T)=P_{\rm TM}(a,T)-P_{\rm TM}^{\rm evan}(a,T),
\label{eq57}
\end{equation}
\end{linenomath}
where $P_{\rm TM}$ is already computed by the Lifshitz formula (\ref{eq53})
written in terms of the Matsubara frequencies.

The computational results for $P_{\rm TM}$, $P_{\rm TM}^{\rm evan}$, and
$P_{\rm TM}^{\rm prop}$ at $T=300~$K normalized to the Casimir pressure between
two ideal metal plates in the classical limit~\cite{10}

\begin{figure}[H]
\includegraphics[width=5.3in]{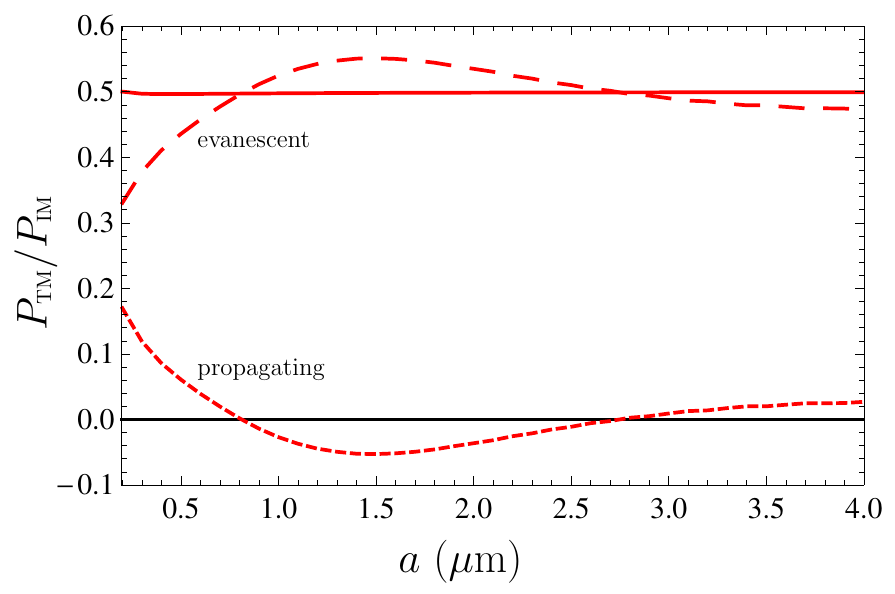}
\caption{The contributions of the transverse magnetic
polarizations to the Casimir pressure between two graphene sheets at $T=300~$K
and to its parts determined by the evanescent and propagating waves normalized
to the Casimir pressure between two ideal metal plates in the classical limit
are shown  as the function of separation by the solid, long-dashed, and
short-dashed lines, respectively. }
\label{fg5}
\end{figure}

\begin{linenomath}
\begin{equation}
P_{\rm IM}(a,T)=-\frac{k_BT}{4\pi a^3}\,\zeta(3),
\label{eq58}
\end{equation}
\end{linenomath}
where $\zeta(z)$ is the Riemann zeta function, are presented in Figure~\ref{fg5}
as the function of separation by the solid, long-dashed, and short-dashed lines,
respectively.

According to Figure~\ref{fg5}, at separations of 200--400~nm both the
evanescent and propagating transverse
magnetic waves contribute significantly to the
Casimir pressure. At larger separations, the dominant contribution is given by
the evanescent waves. In doing so, the contribution of evanescent waves is
attractive at all separation distances. Calculations show, however, that this
attraction is combined from the attractive part caused by the plasmonic region
(\ref{eq42}) and the repulsive part caused by the region (\ref{eq43}).
The contribution of the TM propagating waves to the Casimir pressure between
two graphene sheets changes its character from attraction to repulsion and
vice versa with increasing separation.

\section{Discussion: Whether Graphene Helps to Solve the Problem
Arising for Real Metals} \label{sec5}

The main distinctive feature of the Casimir pressure in the configuration
of two graphene sheets considered above is that in the framework
of the Dirac model, the spatially nonlocal dielectric permittivities
of graphene are found precisely starting from the first principles of
thermal quantum field theory. As to the dielectric permittivities
of metals used in computations by means of the Lifshitz formula,
they contain phenomenological parameters, such as the relaxation
parameter of the Drude model, and have not been tested
experimentally within all frequency regions essential for the Casimir
effect (i.e., in the region of transverse electric evanescent waves).

\textls[-25]{The formalism of the Lifshitz theory for two graphene sheets presented
in \mbox{Sections \ref{sec2} and \ref{sec3}}} is in perfect analogy with that commonly used for
two metallic plates. The Lifshitz formula for the Casimir pressure
remains unchanged, and only the 3D Fresnel reflection coefficients are
replaced with their 2D analogues as it should be done when considering
the Casimir interaction of plane structures. Taking into account the
fundamental character of the Lifshitz theory, we obtain the conclusion
that only some drawback in the used response functions of metals to
the electromagnetic field could cause a disagreement of the theoretical
predictions with measurements of the Casimir interaction between Au
surfaces.

As shown in Section \ref{sec4}, for two graphene sheets, the total Casimir
pressure is determined by the contribution of only the transverse
magnetic waves. This is because in the application region of the Dirac
model at $a\geqslant 200~$nm the Casimir force between graphene sheets is
already in the classical limit where the contributions of the TE
polarized propagating and evanescent waves cancel each other. The
same occurs for the Casimir force between metallic plates described by
the Drude model at separations exceeding the thermal length~\cite{58},
i.e., larger than $7.6~\upmu$m at room temperature. At 
so large separations, however, 
there are 
no reliable measurement data available. As to the
experimental separations between metallic plates, both the TM and TE
polarizations contribute to the Casimir pressure 
irrespective of
whether the
experimentally consistent plasma model or the Drude model excluded by the measurement
data 
is used~\cite{15}.

By and large, the case of graphene suggests to us that when calculating
the Casimir force using the Lifshitz theory, it is important to
adequately describe the response of boundary materials to both the
propagating and evanescent waves with the transverse magnetic and
transverse electric polarizations and take proper account of the
effects of spatial dispersion.

\section{Conclusions} \label{sec6}

In the foregoing, we considered the Casimir pressure between two
graphene sheets using the Lifshitz theory in the form that is most
frequently used for a description of the Casimir effect between
conventional 3D materials. For this purpose, we presented the detailed
derivation of the 2D Fresnel reflection coefficients on a graphene
sheet with due account of the spatial dispersion. As a result, the
reflection coefficients for two independent polarizations of the
electromagnetic field were expressed via the transverse and longitudinal
dielectric permittivities of graphene, which depend on the 2D wave vector,
frequency, and temperature. These reflection coefficients are equivalent
to those expressed directly via the polarization tensor of graphene.

Next, we presented the explicit expressions for the transverse and
longitudinal dielectric permittivities of graphene along the real
frequency axis in the regions of both the propagating and evanescent
waves and also at the pure imaginary Matsubara frequencies. This was
made using the polarization tensor of graphene, which was found earlier
in the framework of the Dirac model.

Using the Lifshitz formula written in terms of the Matsubara frequencies,
we demonstrated that the total Casimir pressure between two graphene
sheets at separations exceeding 200~nm is fully determined by the TM
polarized electromagnetic field. By applying the Lifshitz formula along
the real frequency axis, the contributions of the TM polarized propagating
and evanescent waves to the total pressure were found.

Finally, the above results obtained for graphene sheets were confronted
with the corresponding results valid for two metallic plates. This
confrontation points the way for bringing the Lifshitz theory in
agreement with the measurement data by using the more accurate dielectric
functions of metallic test bodies. {{In the future, it is planned to
consider different contributions to the Casimir force between two real
graphene sheets possessing the nonzero chemical potential, which prevents
from reaching the classical limit at the experimental separations.}}

\vspace{6pt}
\funding{{G.L.K. was partially} funded by the
Ministry of Science and Higher Education of the Russian Federation
(``The World-Class Research Center: Advanced Digital Technologies'',
Contract No. 075-15-2022-311, dated 20 April 2022). The research
of V.M.M. was partially carried out in accordance with the Strategic
Academic Leadership Program ``Priority 2030'' of Kazan Federal
University. } 

\begin{adjustwidth}{-\extralength}{0cm}

\reftitle{References}

\end{adjustwidth}
\end{document}